# Unveiling interpretable development-specific gene signatures in the developing human prefrontal cortex with ICGS

Meng Huang[1], Xiucai Ye[1,2,*] and Tetsuya Sakurai[1,2]

[1]Department of Computer Science, University of Tsukuba, Tsukuba, 3058577, Japan, [2]Center for Artificial Intelligence Research, University of Tsukuba, Tsukuba, 3058577, Japan

[*]Corresponding author: Xiucai Ye. Email: yexiucai@cs.tsukuba.ac.jp

## Abstract

The human prefrontal cortex (PFC) is a highly specialized functional region comprising billions of neuronal and non-neuronal cells in human brain and controls social behavior, decision-making, cognition, personality and memory in human. Development-related changes in human PFC may lead to neurodevelopment disorders including Autism Spectrum Disorder (ASD). To understand the neurodevelopmental processes in human PFC, we explore interpretable gene signatures corresponding to development heterogeneity. Nevertheless, most of the existing gene selection methods identify gene signatures based on correlation, and fail to consider causality between genes. In this work, to unveil interpretable development-specific gene signatures in human PFC, we propose a novel gene selection method, named Interpretable Causality Gene Selection (ICGS), which adopts a Bayesian Network (BN) to represent causality between multiple gene variables and a development variable. The proposed ICGS method combines the positive instances-based contrastive learning with a Variational AutoEncoder (VAE) to obtain this optimal BN structure and use a Markov Blanket (MB) to identify gene signatures causally related to the development variable. Moreover, the differential expression genes (DEGs) are used to filter redundant genes before gene selection. In order to identify gene signatures, we apply the proposed ICGS to the human PFC single-cell transcriptomics data. The experimental results demonstrate that the proposed method can effectively identify interpretable development-specific gene signatures in human PFC. Gene ontology enrichment analysis and ASD-related gene analysis show that these identified gene signatures reveal the key biological processes and pathways in human PFC and have more potential for neurodevelopment disorder cure. These gene signatures are expected to bring important implications for understanding PFC development heterogeneity and function in humans.

## Keywords

## 1. Introduction

The highly developed human brain is the most complex organ in the human nervous system and is divided into different functional regions and development stages, which distinguish humans from other mammals [1, 2, 3]. The human prefrontal cortex (PFC) is a highly specialized functional region comprising billions of neuronal and non-neuronal cells in human brain [2, 3] and plays an important role in the essential aspects of social behavior, memory, decision-making, cognition and personality [3, 4, 5, 6] in human. The development of the brain is a highly complex process including molecular changes, precise gene regulation, glucose metabolism and cell differentiation [7, 8, 9]. Development-related changes in human PFC may lead to neurodevelopment disorders, such as Autism Spectrum Disorder (ASD) and Alzheimer's Disease (AD) [1, 3, 7]. To understand the neurodevelopmental processes in human PFC, it is necessary to explore interpretable gene signatures corresponding to development heterogeneity. Recently, single-cell RNA-sequencing (scRNA-seq) technologies revolutionize the throughput and resolution of bulk RNA sequencing in transcriptomics research [10, 11]. A large amount of single-cell gene expression profiles related to human brain can be obtained due to the appearance of advanced high-throughput sequencing techniques [1, 3]. The advance in human brain single-cell transcriptomics sequencing data has provided a way to investigate development-specific gene signatures related to developing human prefrontal cortex at the single-cell level.

Development-specific gene signature identification of single-cell transcriptomics data is helpful for distinguishing different development stages in human PFC, which can be applied to diverse downstream expression analysis [12, 13, 14]. Many learning methods have been used to identify gene signatures across cell subpopulations or sample groups [15, 16, 17]. For example, Kim et al. and Wang et al. proposed gene selection methods, which utilize the gene profiles with redundant information as gene signatures to classify cancer subtypes [15, 18]. To cut useless genes, lots of feature selection algorithms are used to identify effective gene signatures. For instance, Mundra et al. proposed T-score [19] and Relief-F [20, 21] to view these genes as individual features and identified gene signatures according to top-ranking genes. Guyon et al. combined a common feature selection algorithm (Recursive Feature Elimination, RFE) with Support Vector Machines (SVM), i.e., SVM-RFE [22] to identify gene signatures. To select important genes and filter redundant genes, Reyes et al. applied the extended Relief-F to multi-label learning [23]. Ryan et al. developed a gene selection method (MultiSURF) applied to continuous features for bioinformatics data mining [24, 25]. Other methods use the network-based ranking to identify genes as indicators for biological data [17, 26, 27]. Nevertheless, all the above gene selection methods select genes for each sample based on correlation. Since the correlation is regarded as a statistical basis lacking of interpretability, it is necessary to consider causality between genes [28].

In this work, we considered unveiling interpretable development-specific gene signatures using the gene-related causality between adjacent development stages. A novel gene selection method named interpretable causality gene selection (ICGS) was proposed to select interpretable gene signatures between adjacent developments. We adopted a Bayesian Network (BN) [29, 30, 31, 32] to represent causality between multiple gene variables and a development variable. To obtain this optimal BN structure, the proposed ICGS method combines the positive instances-based contrastive learning [33, 34, 35] with a variational auto-encoder (VAE) [31, 32]. Markov Blanket (MB) [29, 36] is applied to identify genes causally related to the development variable. To better understand human PFC development, we view these selected genes as interpretable development-specific gene signatures for the different development stages of human PFC. Moreover, we use the differently expressed genes (DEGs) to filter redundant genes before applying the proposed method. Different from other gene selection methods, ICGS first infers interpretable development-specific gene signatures using causality between multiple gene variables and a development variable. To demonstrate the effectiveness of ICGS, we apply ICGS into single-cell transcriptomics data of the developing human prefrontal cortex. The experimental results show that the proposed ICGS method can effectively identify interpretable development-specific gene signatures in human PFC and improves the accuracy of PFC development classification according to these selected gene signatures. Gene ontology enrichment analysis illustrates that these identified genes reveal the key biological processes and pathways in human PFC. ASD-related gene analysis shows that these gene signatures have more potential for the neurodevelopment disorder cure. These selected gene signatures are expected to bring important implications for understanding PFC development heterogeneity and function in humans. Here, we firstly illustrate the proposed causality-based ICGS method can select interpretable gene signatures.

## 2. Materials

2.1 Single-cell transcriptomics data in the developing human prefrontal cortex

We obtained single-cell transcriptomics data in the developing human prefrontal cortex from the Gene Expression Omnibus database (the accession number is GSE104276). In this human PFC study, two expected count matrices (Counts and TPM) are provided [3]. Counts are the raw data of gene expression in the transcriptome sequencing, and TPM is the gene expression processed by using the standardization methods. We only used TPM, which contains 2,394 single cells and 24,153 genes. According to the microdissected radial sections of the tissue at gestational weeks (GW), the dataset contains nine gestational weeks (GWs): GW08, GW09, GW10, GW12, GW13, GW16, GW19, GW23, and GW26 [3]. Given the division of development stages in human brain [1], we divided all GWs into six development stages (P2 including GW08 and GW09, P3 including GW10 and GW12, P4 including GW13, P5 including GW16, P6 including GW19 and GW23, and P7 including GW26). The number of single-cells in six human prefrontal cortex development stages (P2, P3, P4, P5, P6 and P7) is 111, 279, 24, 789, 444, and 747, respectively.

Since we need to explore interpretable gene signatures between adjacent development stages, we only use three development stages (P5, P6 and P7) in this human PFC dataset. As a result, we have obtained 24,153 genes and 789 single-cells in P5, 24,153 genes and 444 single-cells in P6, and 24,153 genes and 747 single-cells in P7 from the developing human prefrontal cortex.

2.2 Data pre-processing

There are 24,153 genes and 1,980 single-cells in three development stages (P5, P6 and P7). Firstly, we remained ubiquitous genes and filtered rare genes expressed in less than 10 cells [37], which is less useful for exploring intrinsic transcriptomic signatures of brain cells in the human PFC scRNA-seq data. The remaining genes are helpful for identifying different cell subpopulations (development stages) in human PFC. Besides, we used $\log_e(x+1)$ transformation to further normalize the gene expression data. After gene filtering, we obtained 18,021 genes in P5, P6 and P7. Next, we utilized the EMDomics tool [38] to identify DEGs between adjacent development stages in human PFC single-cell transcriptomics data. These selected DEGs are also more helpful for identifying development stages in human PFC. Finally, we obtained 101 DEGs between P5 with 789 single-cells and P6 with 444 single-cells, and 93 DEGs between P6 with 444 single-cells and P7 with 747 single-cells.

# 3. Methods

### 3.1 Notations

The single-cell gene expression matrix is denoted as $\hat{X} = [x_1, x_2, x_3, \cdots, x_n]^T \in \mathbb{R}^{n \times (m-1)}$, where $x_k \in R^{(m-1)}$ represents the $k$-th cell, and $n$ and $m$-1 are the number of cells and genes, respectively. The label vector of cells is denoted as $p \in \{0, 1\}^{n \times 1}$ that represents the class of cells between adjacent development stages. We combine $\hat{X}$ with $p$ to obtain a sample $X = [x_1, x_2, x_3, \cdots, x_n]^T \in \mathbb{R}^{n \times m}$. Besides, the m-1 genes can be represented as g₁, g₂, ⋯, g$_{m-1}$. This sample matrix $X$ can also be formalized as $X = [g_1, g_2, g_3, \cdots, g_{m-1}, p]$.

### 3.2 Contrastive Learning

Contrastive learning is a self-supervised learning algorithm and is to learn a representation space, where similar samples are close to each other while different samples are far apart [39]. To solve more complicated learning tasks, researchers proposed many frameworks, such as positive instances-based and negative samples-based contrastive learning [34, 35]. For the positive instances-based contrastive learning, positive samples are constructed by using data augmentation (Gaussian Noise) [40]. For example, a positive sample $X' = [x_1', x_2', x_3', \cdots, x_n']^T \in \mathbb{R}^{n \times (m-1)}$ can be obtained for the gene expression matrix $\hat{X}$. In the positive instances-based contrastive learning, the sample $\hat{X}$ is used to learn an encoder that represents an embedding space $Z$ and the positive sample $X'$ is used to obtain the embedding space $Z'$. To represent differences between $Z$ and $Z'$, researchers proposed multiple contrastive losses [41]. Here, Mean Square Error (MSE) is viewed as contrastive loss for $Z$ and $Z'$. Accordingly, we obtain the following contrastive loss $\|Z - Z'\|_2$.

### 3.3 Linear Structural Equation Model

Structural Equation Model (SEM) is a set of covariance-based statistical techniques, which is to describe the data-generating mechanism for a set of variables and analyse the relationships (causality) of observed and latent variables [42, 43]. Linear SEM can be applied to exploring the linear causal relationships among variables [32, 43]. Here, let $W \in \mathbb{R}^{m \times m}$ be the weighted adjacency matrix (learnable). $X = [g_1, g_2, g_3, \cdots, g_{m-1}, p]^T \in \mathbb{R}^{m \times n}$ can be viewed as a sample of a joint distribution of $m$ variables, which each row corresponds to one variable. The linear SEM model is as following:

$$X = W^{\mathsf{T}}X + H, \tag{1}$$

where $H \in \mathbb{R}^{m \times n}$ is the noise matrix and the matrix $W$ is the upper triangular matrix with the sorted nodes via topological order. Hence, the linear SEM is also equivalent to generating a random noise $H$ as follows:

$$X = (I - W^T)^{-1}H. \tag{2}$$

## 3.4 Bayesian Networks

Bayesian Network (BN) is a probabilistic graphical model representing a set of variables and their conditional dependencies using a directed acyclic graph (DAG) [44, 45]. Here, we define a causality matrix $B$ to represent a BN using the weighted adjacency matrix $W \in \mathbb{R}^{m \times m}$ [44], since the BN is often used to represent causal relationships among variables. The causality matrix is as follows:

$$B = W * W^T, \tag{3}$$

where $W \in \mathbb{R}^{m \times m}$ is the weighted adjacency matrix in linear SEM. In a BN, $B$ is the upper triangular matrix that represents the causality relationship across $m$ nodes and the Markov Blanket (MB) of a node (variable) is a set of nodes including its parent nodes (direct causes), child nodes (direct effects) and any other parent nodes of its child nodes [46, 47].

## 3.5 Variational Autoencoder

Variational Autoencoder (VAE) is a deep generative model using variational bayesian methods, which is parameterized by the artificial neural network and is able to capture complex distributions of data [31, 48, 49]. For the inference model (encoder), the noise matrix $H$ in Eq. 1 can be viewed as an embedding space encoded by the encoder in VAE. To obtain the embedding space $H$, Eq. 1 is transformed as follows:

$$H = (I - W^T)X. \tag{4}$$

Generally,

$$H = f'_W(X), \tag{5}$$

where $f'_W$ is the parameter function performing nonlinear transformation in the encoder of VAE. Besides, $f'_W$ can be approximated by the Multilayer Perceptron (MLP) and Eq. 5 is equivalent to $H = (I - W^T)\text{MLP}(X)$.

For the generative model (decoder), Eq. 2 is generally transformed as follows:

$$X = f''_W(H), \tag{6}$$

where $f''_W$ is the parameter function performing nonlinear transformation in the decoder of VAE. In general, $f''_W$ can be approximated by the MLP and Eq. 6 is equivalent to $X = \text{MLP}((I - W^T)^{-1}H)$. Generally, given a specification of the distribution of embedding space $H$ and a set of samples $\tilde{X} = \{X^1, X^2, \ldots, X^j\}$, one may learn the generative model by maximizing the log-evidence:

$$\frac{1}{j}\sum_{k=1}^{n} \log p(X^k) = \frac{1}{n}\sum_{k=1}^{n} \log \int p(X^k|H)p(H)dH, \tag{7}$$

where the solving process is intractable. Therefore, variational bayes is applied to solve it. To approximate the actual posterior $p(H|X)$, we use a variational posterior $q(H|X)$, which is to obtain the evidence lower bound (ELBO) [31, 49]:

$$L_{ELBO} = \frac{1}{j}\sum_{k=1}^{j} L_{ELBO}^k, \tag{8}$$

$$L_{ELBO}^k = -D_{KL}\Big(q(H|X^k)||p(H)\Big) + E_{q(H|X^k)}[\log p(X^k|H)], \tag{9}$$

where $D_{KL}$ is KL-divergence between the variational posterior and the actual one. Besides, the probability density function $q(H|X^k)$ can be used to encode the embedding space $H$ in the inference model of VAE and the probability density function $p(X^k|H)$ can be used to reconstruct $X^k$ in the generative model of VAE. $q(H|X^k)$ and $p(X^k|H)$ can be parameterized by the Neural network (MLP). Here, both $X^k$ and $H$ are $m \times m$ matrices like $X$.

Specifically, the output of inference model in VAE [31, 32] is:

$$[M_H \,|\, \log S_H] = (I - W^T)\text{MLP}(X, P'), \tag{10}$$

where $P'$ is the parameter matrix of MLP, and $M_H$ and $S_H$ are mean value and standard deviation for $H$, respectively. The output of generative model in VAE [31, 32] is:

$$[M_X \,|\, \log S_X] = \text{MLP}((I - W^T)^{-1}H, P''), \tag{11}$$

where $P''$ is the parameter matrix of MLP, and $M_X$ and $S_X$ is mean value and standard deviation for $X$, respectively. Accordingly, based on Eq. 10 and Eq. 11, the corresponding KL-divergence of Eq. 9 is as follows [32]:

$$D_{KL}(q(H|X)||p(H)) = \frac{1}{2}\sum_{i=1}^{m}\sum_{j=1}^{n}(S_H)_{ij}^2 + (M_H)_{ij}^2 - 2\log(S_H)_{ij} - 1. \tag{12}$$

The reconstruction accuracy is obtained by using Monte Carlo approximation as follows [32]:

$$E_{q(H|X)}[\log p(X|H)] = \frac{1}{L}\sum_{l=1}^{L}\sum_{i=1}^{m}\sum_{j=1}^{n} -\frac{x_{ij}-\left(M_X^{(l)}\right)_{ij}^2}{2\left(s_X^{(l)}\right)_{ij}^2} - \log(S_X^{(l)})_{ij} - c, \tag{13}$$

where $M_X^{(l)}$ and $S_X^{(l)}$ are the outputs of the decoder (generative model) in VAE and $c$ is a constant.

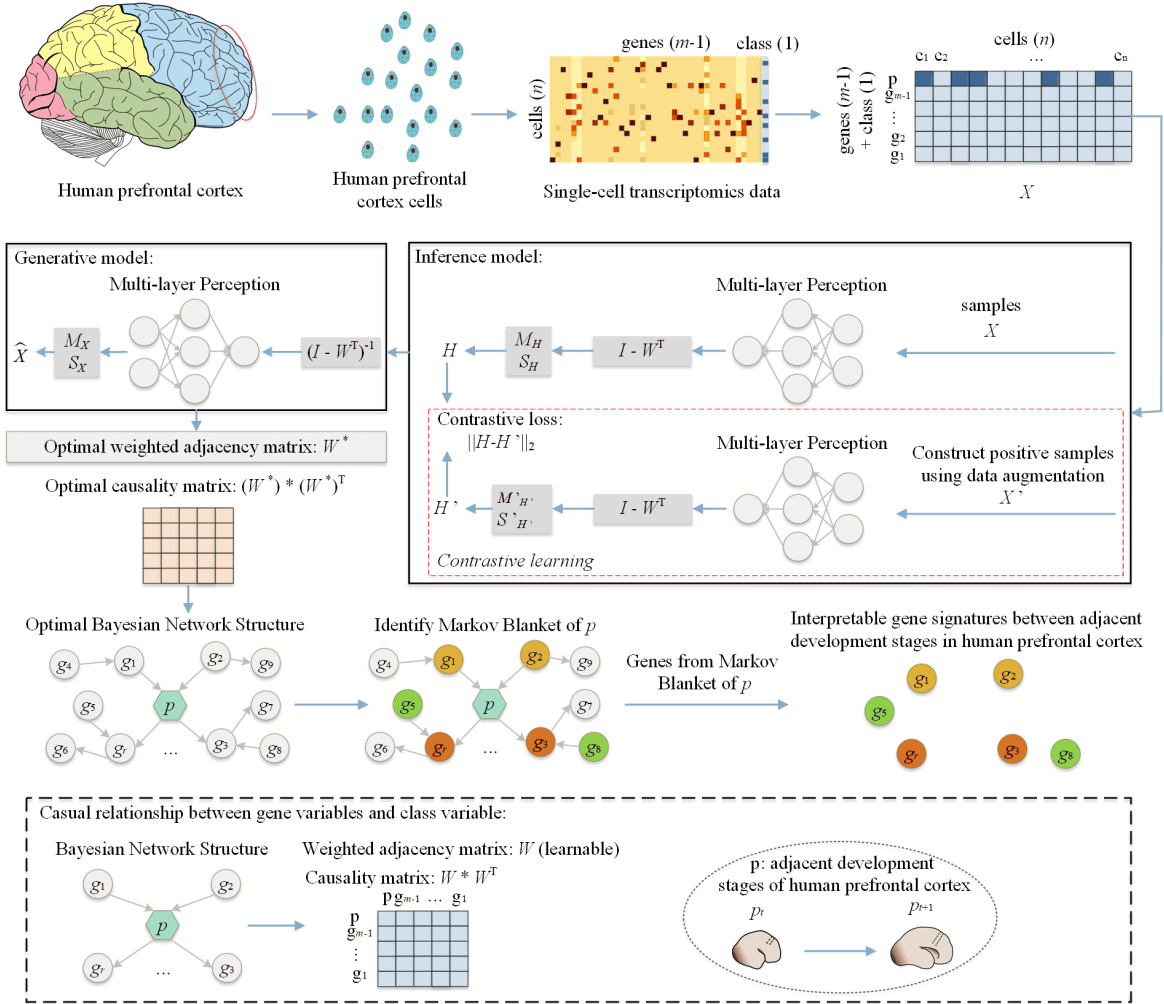

**Figure 1.** Framework of gene signature identification using the proposed method (ICGS) to select interpretable gene signatures between adjacent development stages in the human PFC single-cell transcriptomics data.

3.6 Proposed Method

To identify interpretable gene signatures, we propose a novel interpretable causality gene selection (ICGS) method by combining positive instances-based contrastive learning with VAE. The framework is shown in Figure 1. Firstly, we download a human PFC single-cell transcriptomics dataset, which is preprocessed to obtain the sample $X$ including single-cell gene expression data and the development class. Then, we construct the positive sample $X'$ using data augmentation, which is to enhance the ability of the encoder (inference model) via contrastive learning. Next, we utilize VAE to learn the weighted adjacency matrix $W$, which is to obtain the causality matrix $B = W * W^T$ that represents causality across $m$ nodes including $m$-1 gene variables and 1 development variable in the Bayesian Network (BN). Finally, we obtain the optimal bayesian network structure (BNS) across $m$ nodes. In the BN, we identify Markov Blanket (MB) of the development variable $p$. A set of genes in MB of $p$ are viewed as interpretable gene signatures between adjacent development stages in human prefrontal cortex.

In the inference model of VAE, we apply the positive samples-based contrastive learning to enhancing the ability of the encoder. Accordingly, we obtain embedding spaces $H$ and $H'$, respectively. We define the contrastive loss $L_{con}$ between $H$ and $H'$ via MSE as follows:

$$L_{con} = \|H - H'\|_2. \tag{14}$$

In this proposed framework, the learning objective of VAE is to minimize the negative ELBO $-L_{ELBO}$ and the contrastive loss $L_{con}$ as follows:

$$\min_{W,\theta} \quad -L_{ELBO} + \alpha L_{con}$$

$$s.t. \ \text{tr}[(I + \gamma \cdot B \circ B)^m] = m, \tag{15}$$

where the hyperparameter $\alpha$ is used to adjust the importance between the negative ELBO and the contrastive loss, $B = W * W^T$, $B \circ B$ is the elementwise square of $B$, tr is the trace of matrix, $m$ is the number of variables including gene variables and a development variable (class), $W$ is the weighted adjacency matrix (learnable), $\theta$ is all parameters of VAE, and $\gamma > 0$. Given that BNS is acyclic, we use the acyclicity constraint for $B$.

## 3.7 Evaluation Metrics

To evaluate the proposed ICGS method, we adopted the accuracy metric for classification of adjacent development stages in human PFC, which are defined as follows:

$$Accuracy = \frac{TP+TN}{TP+TN+FP+FN}, \tag{16}$$

where $TP$, $TN$, $FP$ and $FN$ represent true positives, true negatives, false positives and false negatives, respectively. Furthermore, $TP$ is the number of positive samples classified as positive correctly; *TN* is the number of negative samples classified as negative correctly; *FP* is the number of negative samples classified as positive incorrectly; *FN* is the number of positive samples classified as negative incorrectly.

3.8 Parameter Settings

In the proposed ICGS method, there is an important hyper-parameter $\alpha$ to balance the relative importance between the negative ELBO and the contrastive loss. In the experiment, $\alpha$ is set as 1. To ensure the acyclicity constraint for $B$, $\gamma > 0$. Empirically, $\gamma$ can be set as 1 [31, 32]. After obtaining the optimal causality matrix $B$, we keep higher values between variables in $B$ to construct the BN. Accordingly, we need to set the threshold *thr*. The values greater than *thr* in $B$ represent the final causality relationship between variables in BN. Empirically, *thr* can be set as the mean value of the first row elements in $B$ corresponding to the causality relationship values of the development variable *p*. All experiments of the proposed ICGS method are iteratively run 7000 epochs on the four NVIDIA Tesla Ampere A100-PCIe-40GB GPUs and Ubuntu 18.04 system. In the dataset, the percentage of training set and testing set is 70% and 30%, respectively. All experiments are performed independently 100 times, and the average results are reported.

# 4. Results

4.1 Differential Expression Genes Analysis

To preprocess the single-cell transcriptomics data in human PFC, we used the EMDomics tool to identify the differentially expressed genes (DEGs) between adjacent development stages (P5 and P6, P6 and P7). Figure 2 shows density plots of different genes between P5 and P6. For all filtered genes in the development stages P5 and P6, we obtained all emd scores using the EMDomics tool. As shown in Figure 2(A), the emd score with *INSM2* (gene) is 0, which is the lowest value. The highest emd score is 14.66 for *PPP1R17* (gene) in Figure 2(C). Figure 2(B) shows the density plot of *FABP5* (gene) with 7.57 emd scores. We can find that the higher the emd score, the more distinguishable between group A (human PFC cells in P5) and group B (human PFC cells in P6). Accordingly, we obtained 101 genes with emd scores from 7.57 to 14.66 as DEGs. The heatmap plot of 101 DEGs from human PFC cells between P5 and P6 is shown in Figure S2(A) (Supplementary material 1). Similarly, we obtained 93 DEGs from human PFC cells between P6 and P7. As shown in Figure S1 (Supplementary material 1), the density plot of different genes shows the changes of gene in human PFC between P6 and P7. The heatmap plot of 93 DEGs is shown in Figure S2(B) (Supplementary material 1). All detailed information of DEGs is shown in Supplementary material 2.

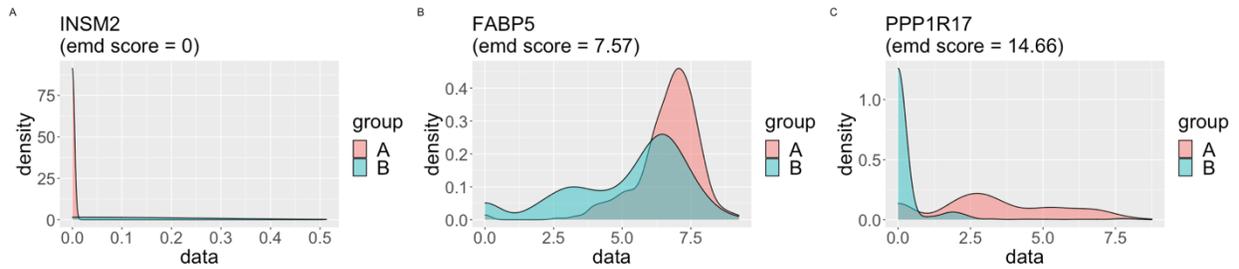

**Figure 2.** The density plot of different genes using the EMDomics tool between adjacent development stages (P5 and P6). (A) The density plot with INSM2 (gene). (B) The density plot with FABP5 (gene). (C) The density plot with PPP1R17 (gene). Group A represents the set of human PFC cells in P5. Group B represents the set of human PFC cells in P6.

## 4.2 Identification of Interpretable Development-specific Gene Signatures

Here, we obtain two bayesian network structures between adjacent development stages (P5 and P6, P6 and P7) using the proposed ICGS. As shown in Figure 3(A), genes variables including *RPS10*, *FBLN1*, *CBS*, *MRPS6*, *CDK1*, *BIRC5*, *FEZF2*, *PRDM8*, *EOMES*, *NEUROD2*, *NPY*, *NUSAP1*, *KRT31* and *IGFBPL1* result in the development variable p (transition of from P5 to P6). In Figure 3(B), genes variables including *ADRBK1*, *MAP6*, *PRKCB*, *PTN*, *NSM*F, *SRCIN1*, *CLU*, *BCAN*, *ATP1A2*, *B4GALNT4*, *HBG1*, *HBA1* and *HBB* result in the development variable p (transition of from P6 to P7), while the development transition p between P6 and P7 leads to *HBA2* and *USP32P1*. In Table 1, we list the rank of identified gene signatures in each human PFC development pair (P5 and P6, P6 and P7). Besides, we also show the heatmap plot of identified gene signatures between adjacent development stages (P5 and P6, P6 and P7) in Figure S3 (Supplementary material 1). In summary, the interpretable development-specific gene signatures between adjacent development stages consist of 14 and 15 genes, respectively.

Given the biological meaning of the selected gene signatures, we performed more detailed analysis. Previous studies that show *NUSAP1* is related to the frontal lobe functions deteriorating resulting in hereditary spastic paraplegias (HSPs) [50], and is helpful for exploring the pathogenesis of neurodegenerative disease [51]. Besides, the frontal cortex research with neurodegenerative diseases shows that *HBA1*, *HBA2* and *HBB* play an important role in variant, iatrogenic forms of Creutzfeldt-Jakob disease (vCJD, iCJD) brains [52] by the significant up-regulation of *HBA1/2* in vCJD brains together with a significant down-regulation of *HBB* in iCJD. Research related to COVID-19 indicates that *HBA1*, *HBA2* and *HBB* are connected to the brain aging-regulated molecular signatures [53]. *EOMES* expresses early neuronal markers in human PFC development [3] and are involved in the neuronal division or migration [54]. Mutations in *EOMES* may lead to malformative microcephaly syndromes [54]. All in all, the identified gene signatures show more significant biological meaning for the study of human brain, especifically PFC development.

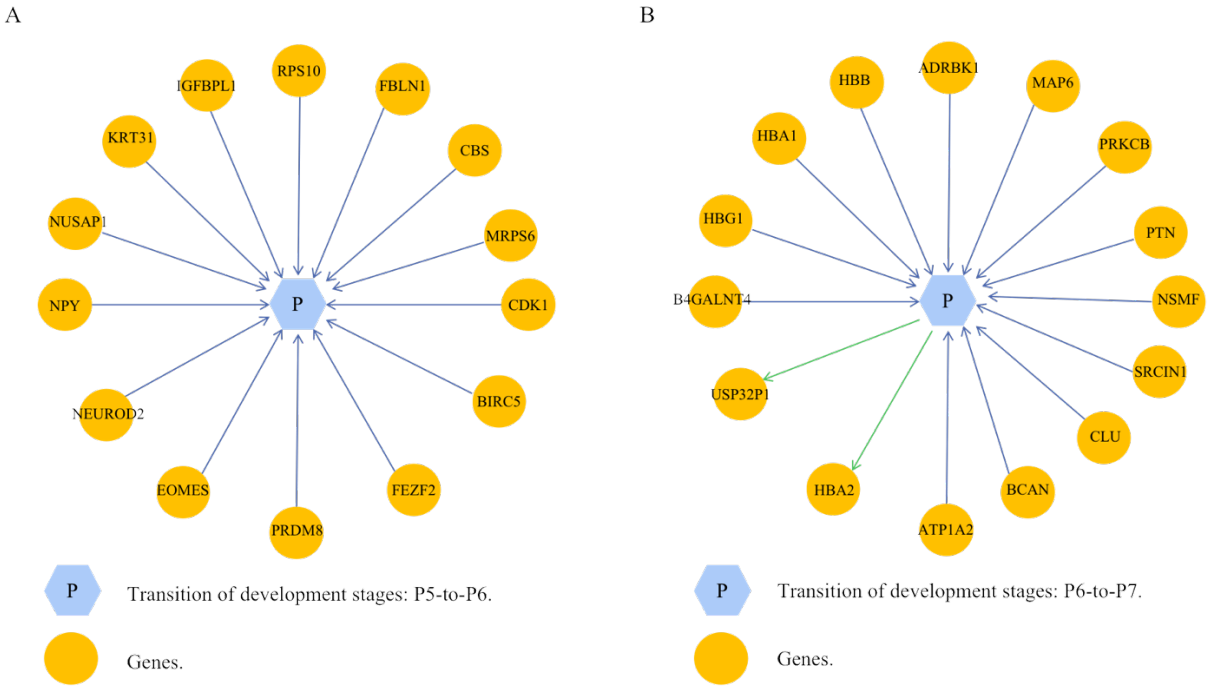

**Figure 3.** The bayesian network structure of identified gene signatures using the proposed ICGS between adjacent development stages (P5 and P6, P6 and P7). (A) P5 and P6. (B) P6 and P7.

**Table 1.** Ranking results of selected gene signatures between adjacent development periods (P5, P6, P7) in the developing human prefrontal cortex.

| Rank | P5toP6 | P6toP7 |
| --- | --- | --- |
| 1 | NUSAP1 | HBA1 |
| 2 | EOMES | HBB |
| 3 | BIRC5 | ATP1A2 |
| 4 | RPS10 | ADRBK1 |
| 5 | CBS | PTN |
| 6 | CDK1 | CLU |
| 7 | NPY | MAP6 |
| 8 | KRT31 | NSMF |
| 9 | FEZF2 | B4GALNT4 |
| 10 | NEUROD2 | SRCIN1 |
| 11 | FBLN1 | PRKCB |
| 12 | IGFBPL1 | HBA2 |
| 13 | PRDM8 | HBG1 |
| 14 | MRPS6 | BCAN |
| 15 | - | USP32P1 |

4.3 Performance Evaluation in Single-cell Transcriptomics Data

To evaluate the effectiveness of identified gene signatures, we compare the proposed ICGS to four state-of-the-art gene selection methods including Chi-square Test (Chi2), Recursive Feature Elimination (RFE) [22], ReliefF [20] and MultiSURF [24] with Random Forest (RF) and Support Vector Machine (SVM) classifiers. To classify adjacent development stages (P5 and P6) in human PFC, the different gene selection methods were performed to rank the genes with the number of selected genes varying from 10 to 14 and we picked the top-ranked genes to train and test different classification models. We compared the classification accuracy of different methods by varying the number of selected genes. As shown in Figure 4(A, B), the classification accuracy gradually increases with increasing gene number for Chi2, RFE, ReliefF, MultiSURF and ICGS with RF and SVM. We can find that the proposed ICGS method has outperformed other methods with the RF and SVM classifiers. Similarly, we compare the classification accuracy of different methods between P6 and P7 for the identified gene signatures with varying the number of genes varying from 11 to 15. As shown in Figure 4(C, D), the classification accuracy of different methods gradually increases with increasing gene number. The proposed ICGS method has higher accuracy than other methods with the number of selected genes varying from 11 to 15. Furthermore, the highest accuracy of RF classifier between adjacent development stages (P5 and P6, P6 and P7) in human PFC was recorded in Table 2 and Table 3 for different gene selection methods. We can find that the proposed ICGS uses selected gene signatures to obtain the highest accuracy 93.4% between P5 and P6 and 89.7% between P6 and P7, respectively. This shows that ICGS has outperformed other state-of-the-art gene selection methods between human PFC adjacent development stages (P5 and P6, P6 and P7). That is because the proposed ICGS selects the interpretable gene signatures corresponding to development stages in human PFC. All in all, these development-specific gene signatures identified by the proposed ICGS method have better distinguishability and more effective information in human PFC development classification.

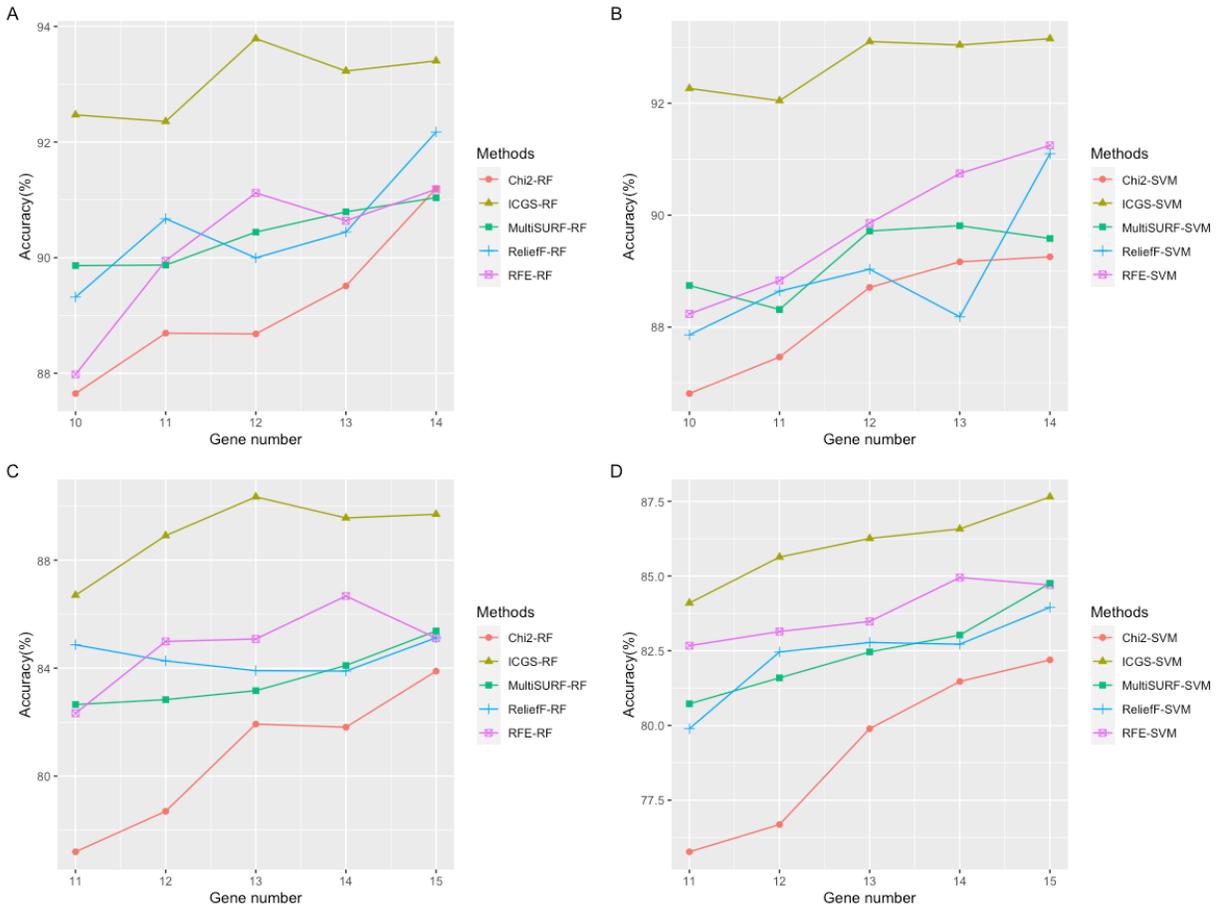

**Figure 4.** Performance evaluation of identified gene signatures between human PFC development stages (P5 and P6, P6 and P7). (A, B) The RF and SVM classification accuracy of proposed ICGS compared with other methods between P5 and P6. The number of selected genes varies from 10 to 14. (C, D) The RF and SVM classification accuracy of proposed ICGS compared with other methods between P6 and P7. The number of selected genes varies from 11 to 15.

**Table 2.** Comparison of different gene identification methods (Chi2, RFE, ReliefF, MultiSURF, and ICGS) to select gene signatures between P5 and P6 (adjacent development periods) by using a RF classifier in the developing human prefrontal cortex.

| Methods | Gene number | | Accuracy |
| | P5 | P6 | |
|---|---|---|---|
| Chi2-RF | 15 | 15 | 91.1 ± 0.61% |
| RFE-RF | 14 | 14 | 91.1 ± 0.40% |
| ReliefF-RF | 14 | 14 | 92.1 ± 0.38% |
| MultiSURF-RF | 14 | 14 | 91.0 ± 0.45% |
| **ICGS-RF** | 14 | 14 | **93.4 ± 0.46%** |

**Table 3.** Comparison of different gene identification methods (Chi2, RFE, ReliefF, MultiSURF, and ICGS) to select gene signatures between P6 and P7 (adjacent development periods) by using a RF classifier in the developing human prefrontal cortex.

| Methods | Gene number | | Accuracy |
|---|---|---|---|
| | P6 | P7 | |
| Chi2-RF | 15 | 15 | 83.8 ± 0.50% |
| RFE-RF | 14 | 14 | 86.6 ± 0.62% |
| ReliefF-RF | 15 | 15 | 85.1 ± 0.70% |
| MultiSURF-RF | 15 | 15 | 85.3 ± 0.72% |
| **ICGS-RF** | 15 | 15 | **89.7 ± 0.48%** |

## 4.4 Functional Enrichment Analysis of Identified Gene Signatures

To study the potential biological functions of gene signatures identified by the proposed ICGS, we perform the enrichment analysis including Gene Ontology (GO), Kyoto Encyclopedia of Genes and Genomes Pathway (KEGG), Reactome, DO biological processes, and DisGeNET. Here, we adopted the R packages GOplot and miRsponge [55] to run the enrichment analysis. The threshold for significance enrichment, *P-value*, is set to 0.01. In Table 4, the top significantly enriched terms are listed for different adjacent development stages (P5 and P6, P6 and P7) in human PFC. All detailed information on Gene Ontology (GO), enrichment KEGG analysis, Reactome, DO biological processes, and DisGeNET are shown in Supplementary material 3. These identified interpretable development-specific gene signatures between adjacent development stages are enriched for biologically important processes that are relevant to human brain including cysteine metabolic process and microcephaly. For example, previous studies demonstrate that cystathionine levels vary greatly between particular human brain regions in the study of the cysteine metabolic process [56]. Besides, the research related to microcephaly shows that the left-right asymmetries of the human cerebral cortex are accompanied by asymmetric gene expression during early fetal development [57, 58]. Overall, these identified development-specific gene signatures may be used to yield important references for finding a solution of human brain development-related disease, illustrating the biological meaning and understanding human PFC development.

**Table 4.** Significant genes and enrichment analysis including the top significantly enriched GO terms, KEGG, Reactome, DO biological processes, and DisGeNET between adjacent development stages (P5 and P6, P6 and P7) in human PFC.

| Adjacent development stages | Gene list | Term type and name | *P-value* |
|---|---|---|---|
| P5 and P6 | NUSAP1 EOMES BIRC5 | GO: cysteine metabolic process | 6.36E-04 |
| | RPS10 CBS CDK1 | KEGG: Glycine, serine and threonine metabolism | 4.88E-03 |
| | NPY KRT31 FEZF2 | Reactome: Sulfur amino acid metabolism | 2.61E-03 |
| | NEUROD2 FBLN1 IGFBPL1 | DO: microcephaly | 3.74E-03 |
| | PRDM8 MRPS6 | DisGeNET: LACTASE PERSISTENCE | 6.92E-04 |
| P6 and P7 | HBA1 HBB ATP1A2 | GO: oxygen transport | 7.95E-04 |
| | ADRBK1 PTN CLU | KEGG: African trypanosomiasis | 4.51E-03 |
| | MAP6 NSMF B4GALNT4 | Reactome: Erythrocytes take up carbon dioxide and release oxygen | 1.21E-03 |
| | SRCIN1 PRKCB HBA2 | DO: macular retinal edema | 1.24E-03 |
| | HBG1 BCAN USP32P1 | DisGeNET: delta beta Thalassemia | 4.61E-04 |

4.5 Autism Spectrum Disorder-related Genes Analysis

Autism spectrum disorder (ASD) is a common neurodevelopmental disorder to describe individuals with a specific combination of impairments in social communication and repetitive behaviours beginning early in life [59, 60]. Previous studies show that 15% of metabolites in human PFC may change their intensities significantly in ASD and development-related changes in human PFC may lead to ASD [59]. Since the human PFC cells are highly correlated with ASD, we obtain ASD-related genes in previous studies [61, 62, 63] and Simons Foundation Autism Research Initiative (SFARI) v2.0 [64] to evaluate the effectiveness of gene signatures identified by the proposed ICGS for ASD. ASD-related genes are shown in Supplementary material 4. We verified whether the identified gene signatures are in the set of ASD-related genes. As shown in Figure S4 (Supplementary material 1), the genes marked in red are the verified ASD-related genes. As a result, we have obtained these identified ASD-related genes including *NPY*, *KRT31*, *FEZF2*, *NEUROD2* and *FBLN1* between P5 and P6, *ATP1A2*, *BCAN*, *CLU*, *PTN* and *PRKCB* between P6 and P7. These genes are helpful for improving ASD patients and providing a solution for ASD cure.

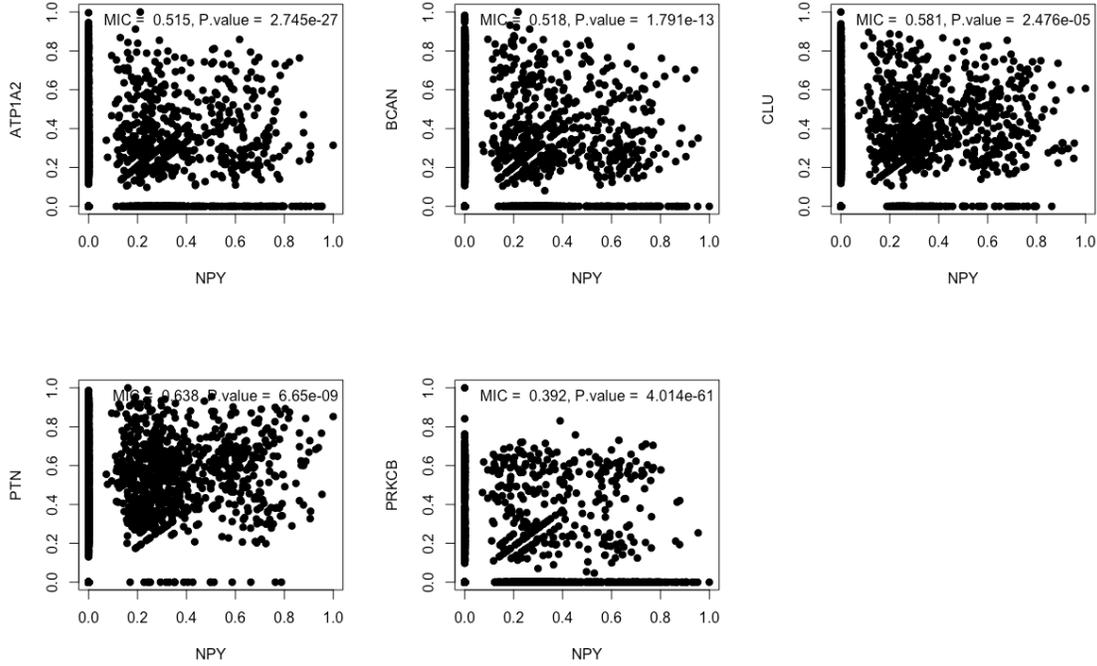

**Figure 5.** Scatter plots of the expression levels between ASD-related gene pairs (*NPY-ATP1A2*, *NPY-BCAN*, *NPY-CLU*, *NPY-PTN* and *NPY-PRKCB*) from the different transition of development stages (P5-to-P6, P6-to-P7). On the top, the corresponding maximal information coefficient (MIC) and *p* value are shown for gene expression values. The two-sided Wilcoxon rank-sum test is used for *p* values. Each dot represents a cell.

Previous studies show that ASD is also a heterogeneous disorder that can reveal a specific genetic disease [65, 66]. Unfortunately, it is difficult to understand the common genetic diseases because gene interactions are non-linear [67, 68]. Unveiling the exact relationships between genes helps to identify regulatory of abnormal gene expression and biological pathways of regulation [69]. To explore the non-linear relationships between identified ASD-related genes from the different transition of development stages (P5-to-P6, P6-to-P7), we use the maximal information coefficient (MIC) [70] to evaluate the non-linear relationships between genes. In statistics, the MIC is a measure of non-linear association between two variables, which belongs to the maximal information-based non-parametric exploration (MINE). Here, we use MIC to measure whether there is a non-linear relationship of ASD-related gene pairs between adjacent development stages (P5 and P6, P6 and P7), such as *NPY* between P5 and P6 and *5 genes* (*ATP1A2*, *BCAN*,

*CLU*, *PTN* and *PRKCB*) between P6 and P7 including *NPY-ATP1A2*, *NPY-BCAN*, *NPY-CLU*, *NPY-PTN* and *NPY-PRKCB*. As shown in Figure 5, we obtain the highest MIC value ($MIC = 0.638, P.value = 6.650 \times 10^{-9}$) for the *NPY-PTN* gene pair and the lowest MIC value ($MIC = 0.392, P.value = 4.014 \times 10^{-61}$) for the *NPY-PRKCB* gene pair. Although the MIC value between ASD-related gene pairs fails to indicate the strong non-linear relationships (MIC > 0.8), there are general and weak non-linear relationships between gene pairs (MIC > 0.3). Similarly, the MIC value of other ASD-related gene pairs is shown in Figure S5, S6, S7 and S8 (Supplementary material 1). We can find that there are still general and weak non-linear relationships between these ASD-related gene pairs. In summary, these identified ASD-related genes are helpful for understanding the disease mechanisms of ASD and have potential to be biomarkers for the solution of ASD cure.

## 5. Conclusions

The human prefrontal cortex (PFC) is very crucial for the cognitive abilities of human beings and is a large part of the neural system that forms normal socio-emotional functions in humans. To understand the neurodevelopmental processes in human PFC, we explore interpretable development-specific gene signatures with the increasing availability of high-throughput sequencing data. To consider causality between genes, we propose an interpretable causality gene selection method to unveil interpretable development-specific gene signatures in human PFC. The causality between gene variables and a development variable is represented by a bayesian network. The positive instances-based contrastive learning and a variational autoencoder are used to obtain this optimal BN structure, where we use the markov blanket of the development variable to causally identify gene signatures. The proposed ICGS is applied to the human PFC single-cell transcriptomics data. The experimental results demonstrate that the proposed method can effectively identify interpretable development-specific gene signatures, which reveal the key biological processes in human PFC and may have the potential for the solution of neurodevelopment disorder (ASD) cure. Although this study has shown the effectiveness of ICGS for the interpretable gene signature identification in human PFC, our method has a wider range of applications in the other biological data. In future studies, we will apply the proposed ICGS method to the data of other human brain regions and the human brain evolution study.

## Data availability statement

The source code used to replicate all our analyses, including all real data, is available at the following link:

https://github.com/linxi159/ICGS.

## Competing interests

The authors declare that they have no known competing financial interests or personal relationships that could have appeared to influence the work reported in this paper.

## Contribution statement

**Meng Huang:** Investigation, Conceptualization, Methodology, Software, Validation, Formal Analysis, Writing - original draft. **Xiucai Ye:** Conceptualization, Methodology, Data curation, Visualization, Resources, Writing - review and editing. **Tetsuya Sakurai:** Supervision, Resources, Project administration, Funding acquisition.

## Supplementary Data

Supplementary data are available online at the following link:

https://github.com/linxi159/ICGS/tree/main/Supplementary_file.

## Acknowledgments

We thank the editor and reviewers for their help and comments during the preparation of the manuscript. We acknowledge the Support for Pioneering Research Initiated by the Next Generation (SPRING) in Japan and the NCBI Gene Expression Omnibus database for providing their platforms and contributors for uploading their meaningful datasets.

## Funding

This work was supported by JST SPRING (JPMJSP2124).